\documentclass[letter,twocolumn]{jpsj2}
\usepackage{amsmath, amsfonts, amsthm, amssymb}
\usepackage{graphicx}
\usepackage{bm}
\usepackage{cite}
\def\runauthor{\textsc{D.~Tahara} and \textsc{M.~Imada}}
\makeatletter
\def\@typeset{}
\makeatother
\def\journal#1#2#3#4{#1 {\bf #2} (#4) #3}
\def\JPSJ{J.\ Phys.\ Soc.\ Jpn.}
\def\PR{Phys.\ Rev.}
\def\PRL{Phys.\ Rev.\ Lett.}
\def\PRB{Phys.\ Rev.\ B}

\def\RMP{Rev.\ Mod.\ Phys.}

\renewcommand{\[}{\begin{equation}}
\renewcommand{\]}{\end{equation}}
\newcommand{\Ha}{\mathcal{H}}
\newcommand{\sP}{\mathcal{P}}
\newcommand{\sL}{\mathcal{L}}

\newcommand{\la}{\langle}
\newcommand{\ra}{\rangle}
\newcommand{\ga}{\alpha}
\newcommand{\gs}{\sigma}
\newcommand{\bdot}{\bm{\cdot}}
\newcommand{\vk}{{\bm{k}}}
\newcommand{\vq}{{\bm{q}}}
\newcommand{\vr}{{\bm{r}}}
\newcommand{\vQ}{{\bm{Q}}}
\newcommand{\Ns}{N_{\text{s}}}
\title{%
Variational Monte Carlo Study of Electron Differentiation around Mott Transition
}
\author{%
Daisuke \textsc{Tahara} and Masatoshi \textsc{Imada}$^{1,2}$
}
\inst{%
Department of Physics, University of Tokyo, 7-3-1 Hongo, Bunkyo-ku, Tokyo 113-0033\\
$^1$Department of Applied Physics, University of Tokyo, 7-3-1 Hongo, Bunkyo-ku, Tokyo 113-8656\\
$^2$JST, CREST, Hongo, Bunkyo-ku, Tokyo 113-8656
}
\abst{%
We study ground-state properties of the two-dimensional Hubbard model at half filling by 
improving  variational Monte Carlo method and by implementing 
quantum-number projection and multi-variable optimization. 
The improved variational wave function enables a highly 
accurate description of the Mott transition and strong fluctuations in metals. 
We clarify how anomalous metals appear near the first-order Mott transition. 
The double occupancy stays nearly constant as a function of the on-site Coulomb interaction 
in the metallic phase near the Mott transition in agreement with the previous unbiased results. 
This unconventional metal at half filling is stabilized by a formation of ``electron-like pockets'' 
coexisting with an arc structure, which leads to a prominent differentiation of electrons 
in momentum space. An abrupt collapse of the ``pocket'' and ``arc'' 
drives the first-order Mott transition. 
}
\kword{%
variational Monte Carlo method, strongly correlated electron systems, Hubbard model, Mott transition, Fermi arc
}
\begin{document}
\maketitle
%
%==============================================================================
When electrons become localized by the strong Coulomb interaction, 
the Mott insulator appears after a metal-insulator transition called the Mott transition \cite{Mott,ImadaFujimoriTokura}. 
The Mott transition is found in many materials and systems, such as transition metal oxides \cite{ImadaFujimoriTokura}, organic conductors \cite{KanodaRev}, and $^3$He layered systems \cite{Ishida,Masutomi}. 
\par
Filling-control Mott transitions at zero temperature have been studied by the auxiliary-field quantum Monte Carlo (AFQMC) method in the Hubbard model on a square lattice \cite{FurukawaImada,FurukawaImada02}. The transition shows a continuous character with divergences of the compressibility and the antiferromagnetic correlation length. 
Bandwidth-control Mott transitions have been studied by the path-integral renormalization group (PIRG) method \cite{Kashima,Morita,Mizusaki02}. The results have shown the first-order as well as 
continuous Mott transitions. 
The Mott transition has also been studied by the dynamical mean-field theory (DMFT) \cite{DMFT02}. The DMFT has shown that the Mott transition at the critical end point at nonzero temperature is consistent with 
the Ising universality class \cite{DMFTIsingUniv}. 
Extensions of the DMFT to include spatial correlations \cite{CPM,CDMFTMIT01,Aichhorn}, such as the cellular dynamical mean-field theory (CDMFT), 
have also suggested the existence of both types of 
the first-order and continuous transitions. 
The Mott transition at finite temperatures has been studied by a phenomenological effective theory \cite{ImadaPRB} and a 
mean-field theory as well \cite{Misawa01,Misawa02}. 
They have found unconventional criticalities at a marginal quantum critical point (MQCP), 
which arises from an interplay of topological transition and symmetry breaking. 
Experimentally measured critical exponents of an organic conductor, $\kappa$-(BEDT-TTF)$_2$Cu[N(CN)$_2$]Cl are in agreement with those of the MQCP \cite{Kagawa}. 
\par
While crucial properties of the Mott transition, such as criticalities, have been 
elucidated by recent extensive studies, 
underlying electronic structure of metals near the Mott insulators has not fully been understood. This is an important issue to be solved, because 
competing orders and quantum phases near the Mott transition 
must be consequences of the underlying unconventional electronic structure. 
Filling-control Mott transitions generating 
the high-temperature superconductivity in the copper oxides, and 
bandwidth-control Mott transitions generating 
the superconductivity and the quantum spin liquid in the organic conductors \cite{Shimizu}
are such consequences. 
\par
%----------------------------------------------------------
Recently, electron differentiations in momentum space around the Mott transition are extensively studied both in experimental \cite{ARPESShen} and theoretical \cite{OnodaImada01,OnodaImada02} studies as a key to understand the unconventional electronic structure. 
The angle resolved photoemission spectroscopy has revealed 
 a truncation of the Fermi surface called Fermi arc, in the underdoped region of the high-temperature copper oxide superconductors (HTSC) \cite{ARPESShen}. 
This truncation is a typical differentiation where electrons at different points of the Fermi surface start showing distinct behavior. 
The origin of the Fermi arc and the pseudogap in the HTSC has been extensively studied by using the extension of DMFT, such as the CDMFT \cite{MPDMFT,ZhangImada}. 
In order to capture the differentiation, an accurate treatment of spatial correlations and high resolution in momentum space are crucially important, while extensions of CDMFT to larger spatial sizes are difficult.
For this purpose, 
the variational Monte Carlo (VMC) method \cite{CeperleyVMC} 
offers an alternative advantageous approach, where 
large system sizes are tractable even with strong interactions and geometrical frustration effects. 
It offers high-resolution results in momentum space. 
However, the bias inherently and inevitably contained in the assumed variational form of the wave functions is a severe 
 limitation in the VMC method. 
Reducing the biases is crucially important in studies with variational wave functions.
\par
In this paper, we study the electron differentiation of the Mott transition by improving the VMC method. 
We apply our improved variational wave function \cite{TaharaVMC} to the two-dimensional 
Hubbard model with geometrical frustration effects. 
We find that highly fluctuating metals with large amplitude of double occupancy of electrons is retained near the Mott transition. 
This unconventional metal at half filling is stabilized by the electron differentiation through a formation of ``electron-like pockets'' together with an arc structure.
%----------------------------------------------------------
\par
The Hubbard model on a square lattice is defined as
\[ \label{eq:Hub}
  \Ha = -\sum_{i,j,\gs} t_{ij} c_{i\gs}^\dag c_{j\gs} + U\sum_{i} n_{i\uparrow} n_{i\downarrow}
,
\]
%%%%%%%
\\[-3mm]
%%%%%%%
where $c_{i\gs}^\dag$ ($c_{i\gs}$) is the creation (annihilation) operator on the $i$-th site with spin $\gs$ and $n_{i\gs} = c_{i\gs}^\dag c_{i\gs}$ is the number operator.
The transfer integral is taken as $t_{ij} =  t$ for the nearest-neighbor sites and $t_{ij} = t'$ for the next nearest-neighbor sites.
We take $\Ns = L\times L$ sites with 
the boundary condition periodic in the $x$ direction and antiperiodic in the $y$ direction (periodic-antiperiodic boundary condition). 
Throughout this paper, we consider the half-filled case $n = (1/\Ns) \sum_{i,\gs} \la n_{i\gs}\ra=1$ at $t'/t=-0.3$.
\par
The variational wave function used in this study has the following form \cite{TaharaVMC}:\\[-5mm]
\[ \label{eq:VWF-fullform}
  |\psi\ra = \sP_{\text{J}}\sP_{\text{d-h}}^{\text{ex.}}\sP_{\text{G}} \sL^{S=0} | \phi_{\text{pair}}\ra
,
\]
where $\sP_{\text{G}}$, $\sP_{\text{d-h}}^{\text{ex.}}$, and $\sP_{\text{J}}$ are 
the Gutzwiller factor \cite{Gutzwiller}, the doublon-holon correlation factor \cite{Kaplan,YokoyamaShiba3}, and the Jastrow factor \cite{Jastrow}, respectively. 
These factors are defined as
\begin{align}
  \sP_{\text{G}} &= \exp \biggl[
    -g\sum_{i} n_{i\uparrow} n_{i\downarrow}
  \biggr]
,
\\[-2mm]
  \sP_{\text{d-h}}^{\text{ex.}} &= \exp\biggl[
   - \sum_{m=0}^{2}\sum_{\ell=1,2} \ga_{(m)}^{(\ell)} \sum_{i} \xi_{i(m)}^{(\ell)}
  \biggr]
,
\\[-1.5mm]
  \sP_{\text{J}} &= \exp \biggl[
    - \frac{1}{2} \sum_{i\neq j} v_{ij} 
      \bigl(n_{i\uparrow}+n_{i\downarrow}\bigr)
      \bigl(n_{j\uparrow}+n_{j\downarrow}\bigr)
  \biggr] 
,
\end{align}
%%%%%%%
\\[-3mm]
%%%%%%%
where $g$, $\ga_{(m)}^{(\ell)}$, and $v_{ij}$ are variational parameters. 
We assume that $v_{ij}=v(\vr_i-\vr_j)$ depends on the displacement $\vr_i-\vr_j$. 
Here, $\xi_{i(m)}^{(\ell)}$ is a many-body operator which is diagonal in the real space representations. 
When a doublon (holon) exists at the $i$-th site and $m$ holons (doublons) surround at the $\ell$-th nearest neighbor, $\xi_{i(m)}^{(\ell)}$ returns $1$. In other cases, $\xi_{i(m)}^{(\ell)}$ returns $0$. 
The spin quantum-number projection $\sL^{S=0}$ restores the $SU(2)$ spin-rotational symmetry and generates a state with total spin $S=0$ \cite{ManyBody,PIRGMizusaki}. 
The one-body part $| \phi_{\text{pair}}\ra$ is the generalized pairing wave function defined as
\begin{align} \notag
  |\phi_{\text{pair}}\ra = 
  \Biggl[ \sum_{\vk\in\text{BZ}}
    &\varphi^{(1)}(\vk)
    c_{\vk\uparrow}^\dag c_{-\vk\downarrow}^\dag	
+\!\!\!\!
    \sum_{\vk\in\text{AFBZ}}
\!\!\!
    \varphi^{(2)}(\vk)
\\[-3mm]
\label{eq:VWF-one-bodyfull}
&
    \times \Bigl( 
        c_{\vk+\vQ\uparrow}^\dag c_{-\vk\downarrow}^\dag
      - c_{\vk\uparrow}^\dag c_{-\vk-\vQ\downarrow}^\dag
    \Bigr)\!
  \Biggr]^{N/2} \!\!\!\!\!| 0 \ra
\end{align}
%%%%%%%
\\[-1mm]
%%%%%%%
with the conditions 
$\varphi^{(1)}(-\vk) = \varphi^{(1)}(\vk)$ and $\varphi^{(2)}(-\vk) = \varphi^{(2)}(\vk)$. 
Here, BZ and AFBZ denote the Brillouin zone and the folded antiferromagnetic (AF) Brillouin zone, respectively. The pair orbitals $\varphi^{(1)}(\vk)$ and $\varphi^{(2)}(\vk)$ are variational parameters. 
Since the PIRG method shows that the ordering vector of the AF insulating phase in the interval, $0\le -t'/t \le 0.5$, is $\vQ=(\pi,\pi)$ \cite{Kashima,Mizusaki02}, 
the vector $\vQ$ in eq. (\ref{eq:VWF-one-bodyfull}) is chosen as $\vQ=(\pi,\pi)$ in this study. 
All the variational parameters ($g$, $\ga_{(m)}^{(\ell)}$, $v_{ij}$, $\varphi^{(1)}(\vk)$, and $\varphi^{(2)}(\vk)$) are simultaneously optimized by using the stochastic reconfiguration method \cite{SorellaSR}. 
The variational wave function $|\psi\ra$ in eq. (\ref{eq:VWF-fullform}) allows to describe various states such as paramagnetic metals (PM), antiferromagnetically ordered states, and superconducting states with any wavenumber (spatial) dependence of gap functions within a single functional form. 
Moreover, $|\psi\ra$ enables efficient treatment of quantum fluctuations with long-ranged as well as short-ranged correlations. 
For example, the peak value of the spin structure factor in the doped Hubbard model shows quantitative agreement with the results obtained from the unbiased AFQMC method. 
Detailed discussions and extensive benchmarks are reported elsewhere \cite{TaharaVMC}. 
\par
%==============================================================================
\begin{figure}[t]
\centering
\includegraphics[width=0.38\textwidth]{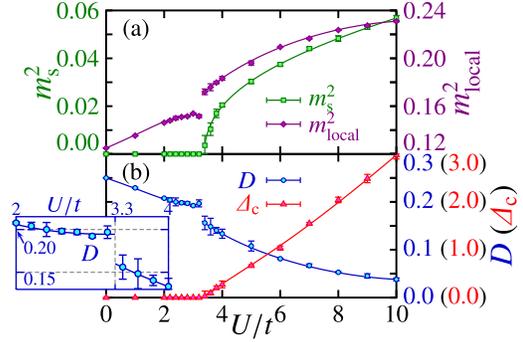}%
\vspace*{-0.6\baselineskip}
\caption{%
(Color online) 
(a) Squared staggered-magnetization $m_{\text{s}}^2$, squared local moment $m_{\text{local}}^2$, (b) double occupancy $D$, and charge gap $\varDelta_{\text{c}}$ in the thermodynamic limit as functions of the on-site interaction $U/t$ at $t'/t=-0.3$. Each solid curves are fitted by the third-order B\'{e}zier curve. 
The inset is an enlargement of (b) for $D$ near the transition. 
}
\label{fig:domscg01}
\vspace*{-1.1\baselineskip}
\end{figure}
%==============================================================================
\par
%----------------------------------------------------------
In order to capture the metal-insulator and magnetic phase transitions, we calculate 
the double occupancy 
$D(\Ns) = (1/\Ns) \sum_{i} \la n_{i\uparrow}n_{i\downarrow}\ra$, 
spin structure factor 
$S(\vq,\Ns) = (1/3\Ns) \sum_{\vq} \la \bm{S}_i\!\bdot\!\bm{S}_j \ra e^{i\vq\bdot (\vr_i-\vr_j)}$, 
and charge gap 
$\varDelta_{\text{c}}(\Ns) = [ \mu( (\Ns+1)/\Ns )-\mu((\Ns-1)/\Ns) ]/2$ 
for $\Ns$-site systems, 
where $\bm{S}_i$ is the spin operator and the chemical potential $\mu$ is given as $\mu((2M-1)/\Ns)=[ E(M,M)-E(M-1,M-1) ]/2$. Here, $E(N_{\uparrow}, N_{\downarrow})$ is the variational energy with the number of up (down) spin $N_{\uparrow}$ ($N_{\downarrow}$). 
The double occupancy in the thermodynamic limit $D$ is estimated 
by fitting the finite-size data in the form 
$
  D - D(\Ns) \propto L^{-1}
$, because the PIRG results for the frustrated Hubbard model have succeeded in the extrapolation by this form and estimating the critical point for the MIT \cite{Kashima,Morita}. 
The staggered magnetization in the thermodynamic limit $m_{\text{s}}$ is also estimated by fitting the scaling form 
$
  m_{\text{s}}^2 - S(\bm{Q},\Ns)/\Ns \propto L^{-1}
$ with $\bm{Q}=(\pi,\pi)$ 
by following the scaling of the spin wave theory if the long-range order is present \cite{Huse}. 
We also calculate the local moment $m_{\text{local}}$ defined as 
$
  m_{\text{local}} = \sqrt{ \la S_i^z S_i^z\ra } = (1/2) \sqrt{n-2D}
$. 
For the charge gap in the thermodynamic limit $\varDelta_{\text{c}}$, we use the scaling function 
$
  \varDelta_{\text{c}} - \varDelta_{\text{c}}(\Ns) \propto L^{-1}
$ as in the AF Hartree-Fock gap equation \cite{FurukawaImada}. 
The extrapolation to the thermodynamic limit is performed by using several choices of system sizes up to $L=16$.
\par
Figure \ref{fig:domscg01} shows the $U/t$ dependence of $D$, $m_{\text{s}}^2$, $m_{\text{local}}^2$, and $\varDelta_{\text{c}}$. 
The first-order metal-insulator transition takes place at $U_{\text{c}}/t=3.3\pm 0.1$. The magnetic phase transition takes place at the same critical point 
within our resolution. 
The nonmagnetic insulating phase is not clearly identified in our variational results. 
Although $m_{\text{local}}$ gradually increases as a function of $U/t$, the growth is strongly suppressed in the metallic phase near the metal-insulator transition. 
In the metallic region, $S(\vQ,\Ns)$ at $\vQ=(\pi,\pi)$ rapidly grows when $U/t$ approaches $U_{\text{c}}/t$, although the true long-ranged order is not achieved within metals.  
This growth is, however, to a large extent, ascribed to the growth in the longer ranged part in the real space correlation, while the shorter ranged part of antiferromagnetic correlations does not grow equally.  In this overall tendency, the local moment $m_{\text{local}}^2$, that is the shortest-ranged part, shows nearly flat dependence on $U/t$. This seems to optimize and reconcile in gaining the kinetic energy by keeping the electron-pocket-like and arc-like structure of the Fermi surface as we describe below. In other words, the flat $U/t$ dependence of $m_{\text{local}}^2 = \lim_{\Ns \to \infty} (1/\Ns) \sum_{\vq} S(\vq,\Ns)$
with growing $S(\vQ,\Ns)$ means a compensating reduction of $S(\vq,\Ns)$ at $\vq \neq \vQ$. This reduction suppresses the electron scattering by the spin fluctuations and keeps the coherence of the carrier, leading to the gain in the kinetic energy. 
\par
Figure \ref{fig:domscg01} shows remarkable quantitative agreement 
with the unbiased results obtained from the PIRG method \cite{Kashima}. 
Although the value $t'/t$ in the PIRG data in ref. \citen{Kashima} is not the same as that of the present results, 
the values at $t'/t=-0.3$ are rather precisely interpolated from the
results at $t'/t=-0.2$ and $t'/t=-0.5$ given by the PIRG method.  For example, 
$U_{\text{c}}/t$ estimated by interpolation of the PIRG results suggests
$U_{\text{c}}/t \sim 3.6 \pm 0.1$, which may be compared with the present estimate
$U_{\text{c}}/t = 3.3 \pm 0.1$.  
The absence or only tiny interval of the nonmagnetic insulating region at $t'/t=-0.3$ is also consistent each other. 
Furthermore, $D$ near $U_{\text{c}}/t$ in the metallic
phase stays nearly constant around $D \sim 0.2$ over an interval $2 \le U/t \le 3.3$ 
contrary to a naive expectation.
This flat $U/t$ dependence of $D$ also remarkably agrees with the PIRG results.
In addition, Fig. \ref{fig:domscg01} shows convex growth of the squared staggered-magnetization $m_{\text{s}}^{2}$ from $U_{\text{c}}/t$ for $U/t > U_{\text{c}}/t$. 
These features have been observed by the PIRG method at $t'/t=-0.2$ as well. 
In our results, the squared staggered-magnetization at $U/t=4.0$ and $t'/t=-0.3$ is $m_{\text{s}}^{2}=0.020\pm 0.002$. In the PIRG results, the value at $U/t=4.0$ and $t'/t=-0.2$ is $m_{\text{s}}^{2}\simeq0.025$. 
When we consider the difference of $t'/t$, this is a remarkable quantitative agreement. 
Our results suggest that our variational wave function enables 
quantitatively accurate descriptions in the ground state. 
\par
On the other hand, in the previous VMC calculation for this model, the critical value of 
the Mott transition is much larger ($U_{\text{c}}/t \sim 6.7$) \cite{YokoVMCdiag}. 
In addition, 
the double occupancy in ref. \citen{YokoVMCdiag} linearly decays to the transition point 
as a function of $U/t$ with a substantial slope ($\partial D/\partial (U/t) \sim -0.033$).
The variational wave functions employed in the literature 
include many-body correlations only in much restricted forms, such as the short-ranged doublon-holon factor.
Such a restricted form does not sufficiently take into account 
fluctuations, which are strongly enhanced around the Mott transition. 
By introducing a large number of variational parameters 
that scales linearly with the system size, 
the Gutzwiller-Jastrow factor as well as the one-body part allow much more accurate treatment of fluctuations . 
DMFT results also show large $U_{\text{c}}/t \sim 11$ even for $t'/t=0$ and fail in capturing the plateau of $D$ at $U/t \le U_{\text{c}}/t$ \cite{DMFT02}. 
Although the CDMFT takes into account spatial correlations to some extent, 
the plateau of $D$ around $U_{\text{c}}/t$ has not been captured yet in the CDMFT studies \cite{CDMFTTri01,CDMFTTri02}. 
The exact diagonalization study \cite{Koretsune} does not capture this behavior either 
because of the limitation of the system size. 
Sufficiently large system size over the correlation length of the fluctuations is important for reproducing the sufficient coherence and the plateau of $D$.
\par
%==============================================================================
\begin{figure}[t]
\centering
\includegraphics[width=0.34\textwidth]{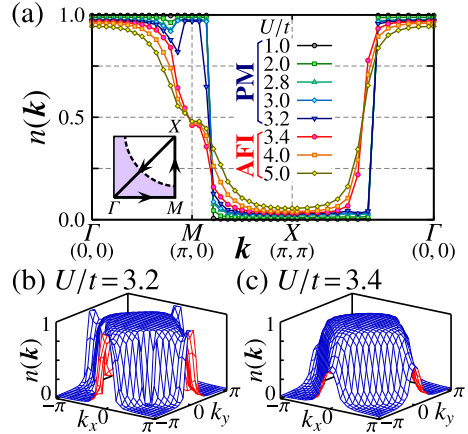}%
\vspace*{-0.36\baselineskip}
\caption{%
(Color online) (a) 
$U/t$ dependence of momentum distribution $n(\vk)$ for $L=14$ and $t'/t=-0.3$. 
Error bars are comparable to the symbol sizes.
The filled points are values of $n(\vk)$ in eq. (\ref{eq:k-pointmesh}) and $n(\vk)$ obtained by the linear interpolation of two nearest $\vk$-points along the vertical direction of each $\varGamma$-$M$-$X$-$\varGamma$ line. PM and AFI denote the paramagnetic metal and the antiferromagnetic insulator, respectively. 
Dashed curve in the inset is the Fermi surface for $t'/t=-0.3$ and $U/t=0$. 
The three-dimensional plot of $n(\vk)$ at (b) $U/t=3.2$ and (c) $U/t=3.4$ for $L=14$ and $t'/t=-0.3$. 
The value $n(\vk)$ at 
$
  \{
  \vk = ( (2\ell_x-1) \pi/L, (2\ell_y-1) \pi/L ), ( 2\ell_x \pi/L, 2\ell_y\pi/L ) \, 
  |\, 
  \ell_x,\ell_y = -L+1, -L+2, \cdots, L-1, L
  \}
$
is obtained from the linear interpolation of $n(\vk)$ in eq. (\ref{eq:k-pointmesh}).
}
\label{fig:nkslice}
\vspace*{-0.91\baselineskip}
\end{figure}
%==============================================================================
\par
In order to analyze the electron differentiation in this region, 
we calculate the momentum distribution 
$
  n(\vk) = ({1}/{2\Ns}) \sum_{i,j,\gs} \la c_{i\gs}^\dag c_{j\gs} \ra e^{i\vk\bdot(\vr_i-\vr_j)}
$. 
We assume that $n(\vk)$ has fourfold symmetric structure. 
Since we calculate $n(\vk)$ under the periodic-antiperiodic boundary condition, 
this assumption allows twice as many $\vk$-points in the BZ for the $L\times L$ system:\\[-5mm]
\[ \label{eq:k-pointmesh}
  \vk = \Bigl( (2\ell_x-1) \frac{\pi}{L}, 2\ell_y \frac{\pi}{L} \Bigr), \Bigl( 2\ell_x \frac{\pi}{L}, (2\ell_y-1)\frac{\pi}{L} \Bigr)
\] \\[-4mm]
for $ \ell_x,\ell_y = -L+1, -L+2, \cdots, L-1, L$. 
The VMC results for $L=14$ are shown in Fig. \ref{fig:nkslice}. 
The momentum distribution $n(\vk)$ shows a characteristic behavior near the Mott transition. 
In the metallic phase, the amplitude of $n(\vk)$ around $M$ points ($\vk = (\pm \pi,0), (0,\pm\pi)$) 
is kept large, generating an ``electron pocket-like'' structure just before the Mott transition as is seen in Figs.~\ref{fig:nkslice}(a)~and~(b). 
%----------------------------------------------------------
\par
\begin{figure}[t]
\centering
\includegraphics[width=0.34\textwidth]{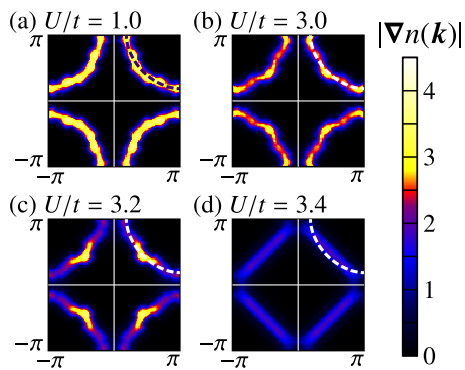}%
\vspace*{-0.4\baselineskip}
\caption{%
(Color online) Contour plots of $|\bm{\nabla}n(\vk)|$ for $L=14$ and $t'/t=-0.3$. 
Here, $|\bm{\nabla}n(\vk)|$ is calculated by using the value from the bicubic interpolation of $n(\vk)$ in eq. (\ref{eq:k-pointmesh}). Dashed curves are the Fermi surface for $t'/t=-0.3$ and $U/t=0$.
}
\label{fig:nkgrad}
\vspace*{-1.3\baselineskip}
\end{figure}

%----------------------------------------------------------
Figures \ref{fig:nkgrad}(a)-(d) are contour plots of $|\bm{\nabla}n(\vk)|$. We can estimate the Fermi surface and the coherence of electrons on the Fermi surface by the amplitude of $|\bm{\nabla}n(\vk)|$. 
The contour plots remarkably show the ``arc-like'' structure in the metallic phase close to the Mott transition (Fig. \ref{fig:nkgrad}(c)). 
Although our system size is not sufficient, 
the position of the Fermi surface looks rigid and very similar to that of the non-interacting system, 
contrary to the ``deformation to the nesting'' obtained in the renormalization-group method in the weak-coupling regime \cite{RGMiyake}. 
The rigidity enhances the coexistence of ``pocket'' and ``arc.''
The ``arc'' around $(\pm\pi/2,\pm\pi/2)$ 
coexisting with the ``electron pocket-like'' structure around $(\pm \pi,0)$ and $(0,\pm\pi)$ 
discussed before is retained near the Mott transition. 
The coexistence implies that, although the long-range magnetic order is absent, the gap structure of the charge excitation already gets formed even in the metallic phase as a precursor of the Mott gap. 
As in semimetals, the preformed Mott gap generates electron and hole pockets \cite{Misawa02}. 
The arc structure emerges as a part of the hole pockets 
around $(\pm\pi/2,\pm\pi/2)$ where outer half of the pocket is lost presumably because of a strong damping. 
The coexistence of the ``electron pocket'' and ``arc'' directly causes largely retained $D$ and suppressed $m_{\text{local}}$ in Fig. \ref{fig:domscg01}. 
This ``semimetallic pocket'' structure creates electrons and holes in the upper and lower Hubbard bands, respectively. As a result, excitons (or doublon-holon pairs) are generated in the background of the Mott insulator, where $D$ is largely retained. 
The first-order Mott transition appears as a sudden collapse of the ``semimetallic pocket'' structure. 
The similar abrupt change of $n(\vk)$ is also seen in the PIRG results \cite{Kashima}. 
The Mott transition emerging as shrinkage of electron-hole (or doublon-holon) pockets supports the topological character of the transition proposed in ref. \citen{Misawa02}.
\par
%==============================================================================
%==============================================================================
\par
In summary, we have studied the electron differentiation in momentum space around the Mott transition by using the improved VMC method. Our variational results show quantitative agreement with the unbiased method. Especially, our variational wave function enables to treat short-ranged as well as long-ranged spin/charge fluctuations, which are strongly enhanced near the Mott transition. 
As a result, we have captured the flat $U/t$ dependence of double occupancy $D$ and abrupt change of momentum distribution $n(\vk)$ at the first-order Mott transition. 
The momentum distribution shows ``electron pocket-like'' structure around $\vk=(\pm\pi,0),(0,\pm\pi)$. 
The arc structure appears around $\vk=(\pm\pi/2,\pm\pi/2)$ in $|\bm{\nabla}n(\vk)|$ in the metallic phase near the Mott transition. 
The abrupt collapse of these parts drives the first-order Mott transition. 
The coexistence of ``electron pocket-like'' structure and ``hole-like arcs'' with the retained plateau of the double occupancy $D$ is the key aspects of the electron differentiation in momentum space. 
Clarifying the relation between the above differentiation and instability toward superconductivity is one of the most important issues left for future studies. 
\par
One of the authors (D.T.) thanks  S.~Watanabe, T.~Misawa, and Y.~Yamaji for useful discussions. 
This work was supported by Grants-in-Aid for
Scientific Research on Priority Areas under the grant numbers 
17071003, 16076212, and 17064004 
from the Ministry of Education, Culture, Sports, Science and
Technology. A part of our computation has been done using
the facilities of the Supercomputer Center, Institute for Solid
State Physics, University of Tokyo.
%==============================================================================

%------------------------------------------------------------------------------

\vspace*{-5mm}
\begin{thebibliography}{99}

\bibitem{Mott}
N. F. Mott and R. Peierls: \journal{Proc.\ Soc.\ London.}{49}{72}{1937}.

\bibitem{ImadaFujimoriTokura} 
For a review see M. Imada, A. Fujimori, and Y. Tokura: \journal{\RMP}{70}{1039}{1998}.

\bibitem{KanodaRev}
K. Kanoda: \journal{\JPSJ}{75}{051007}{2006}.

\bibitem{Ishida} 
K. Ishida, M. Morishita, K. Yawata, and H. Fukuyama: \journal{\PRL}{79}{3451}{1997}.

\bibitem{Masutomi} 
R. Masutomi, Y. Karaki, and H. Ishimoto: \journal{\PRL}{92}{025301}{2004}.

\bibitem{FurukawaImada}
N. Furukawa and M. Imada: \journal{\JPSJ}{61}{3331}{1992}.

\bibitem{FurukawaImada02}
N. Furukawa and M. Imada: \journal{\JPSJ}{62}{2557}{1993}.

\bibitem{Kashima}
T. Kashima and M. Imada: \journal{\JPSJ}{70}{3052}{2001}.

\bibitem{Morita}
H. Morita, S. Watanabe, and M. Imada: \journal{\JPSJ}{71}{2109}{2002}.

\bibitem{Mizusaki02}
T. Mizusaki and M. Imada: \journal{\PRB}{74}{014421}{2006}.

\bibitem{DMFT02}
A. Georges, G. Kotliar, W. Krauth, and M. J. Rozenberg: \journal{\RMP}{68}{13}{1996}.

\bibitem{DMFTIsingUniv}
G. Kotliar, E. Lange, and M. J. Rozenberg: \journal{\PRL}{84}{5180}{2000}.

\bibitem{CPM}
S. Onoda and M. Imada: \journal{\PRB}{67}{161102}{2003}.

\bibitem{CDMFTMIT01}
O. Parcollet, G. Biroli, and G. Kotliar: \journal{\PRL}{92}{226402}{2004}.

\bibitem{Aichhorn}
M. Aichhorn, E. Arrigoni, M. Potthoff, and W. Hanke: \journal{\PRB}{76}{224509}{2007}.

\bibitem{ImadaPRB}
M. Imada: \journal{\PRB}{72}{075113}{2005}.

\bibitem{Misawa01}
T. Misawa, Y. Yamaji, and M. Imada: \journal{\JPSJ}{75}{083705}{2006}.

\bibitem{Misawa02}
T. Misawa and M. Imada: \journal{\PRB}{75}{115121}{2007}.

\bibitem{Kagawa}
F. Kagawa, K. Miyagawa, and K. Kanoda: Nature \textbf{436} (2005) 534.

\bibitem{Shimizu} 
Y. Shimizu, K. Miyagawa, K. Kanoda, M. Maesato, and G. Saito: \journal{\PRL}{91}{107001}{2003}.

\bibitem{ARPESShen}
A. Damascelli, Z. Hussain, and Z.-X. Shen: \journal{\RMP}{75}{473}{2003}.

\bibitem{OnodaImada01}
S. Onoda and M. Imada: \journal{J.\ Phys.\ Chem.\ Solids}{62}{47}{2001}.

\bibitem{OnodaImada02}
S. Onoda and M. Imada: J.\ Phys.\ Chem.\ Solids \textbf{63} (2002)~2225.

\bibitem{MPDMFT}
T. D. Stanescu and G. Kotliar: \journal{\PRB}{74}{125110}{2006}.

\bibitem{ZhangImada}
Y. Z. Zhang and M. Imada: \journal{\PRB}{76}{045108}{2007}.

\bibitem{CeperleyVMC} 
D. Ceperley, G. V. Chester, and M. H. Kalos: \journal{\PRB}{16}{3081}{1977}.

\bibitem{TaharaVMC}
D. Tahara and M. Imada: arXiv:0805.4457.

\bibitem{Gutzwiller}
M. C. Gutzwiller: \journal{\PRL}{10}{159}{1963}.

\bibitem{Kaplan}
T. A. Kaplan, P. Horsch, and P. Fulde: \journal{\PRL}{49}{889}{1982}.

\bibitem{YokoyamaShiba3}
H. Yokoyama and H. Shiba: \journal{\JPSJ}{59}{3669}{1990}.

\bibitem{Jastrow}
R. Jastrow: \journal{\PR}{98}{1479}{1955}.

\bibitem{ManyBody}
P. Ring and P. Schuck: {\it The Nuclear Many-Body Problem}, (Springer-Verlag, New York, Heidelberg, Berlin, 1980).

\bibitem{PIRGMizusaki}
T. Mizusaki and M. Imada: \journal{\PRB}{69}{125110}{2004}.

\bibitem{SorellaSR}
S. Sorella: \journal{\PRB}{64}{024512}{2001}.

\bibitem{Huse}
D. A. Huse: \journal{\PRB}{37}{2380}{1988}.

\bibitem{YokoVMCdiag}
H. Yokoyama, M. Ogata, and Y. Tanaka: \journal{\JPSJ}{75}{114706}{2006}.

\bibitem{CDMFTTri01}
B. Kyung and A.-M. S. Tremblay: \journal{\PRL}{97}{046402}{2006}.

\bibitem{CDMFTTri02}
T. Ohashi, T. Momoi, H. Tsunetsugu, and N. Kawakami: \journal{\PRL}{100}{076402}{2008}.

\bibitem{Koretsune}
T. Koretsune, Y. Motome, and A. Furusaki: \journal{\JPSJ}{76}{074719}{2007}.

\bibitem{RGMiyake}
T. Ogawa, H. Maebashi, H. Kohno, and K. Miyake: \journal{Physica B}{312-313}{525}{2002}.

\end{thebibliography}
\end{document}